\documentclass[12pt,titlepage]{article}

\usepackage{pslatex}
\usepackage[T1]{fontenc}
\usepackage[latin1]{inputenc}

\usepackage{times}
\usepackage{latexsym}
\usepackage{subfigure}
\usepackage{xspace}
\usepackage{ifpdf}
\usepackage{graphicx}
\ifpdf
\fi
\begin{document}

  \title{Analyzing Large Collections of\\Electronic Text Using OLAP\\
         {\small UNBSJ CSAS Technical Report TR-05-001}\\[1cm]}

  \author{Steven Keith, Owen Kaser\\
   University of New Brunswick\\
  Saint John, Canada\\
  {\tt steven.keith@unb.ca, owen@unbsj.ca} \and
  Daniel Lemire\\
  Université du Québec à Montréal (UQAM)\\
  Montréal, Canada\\
  {\tt lemire@ondelette.com}
  }
  \date{June 2005}
\maketitle

\pagestyle{plain}

\begin{abstract}
Computer-assisted reading and analysis of text has applications in the
humanities and social sciences.  Ever-larger electronic text archives
have the advantage of allowing a more complete analysis but the
disadvantage of forcing longer waits for results.  On-Line Analytical
Processing (OLAP) allows quick analysis of multidimensional data.  By
storing text-analysis information in an OLAP system, queries may be
solved in seconds instead of minutes or hours.  This analysis is
user-driven, allowing users the freedom to pursue their own directions
of research.
\end{abstract}
\section{Introduction}
Electronic text collections have existed for over half a century.  In
this time these archives have increased in both size and accuracy.
Many tools have been created for searching, classifying, and
retrieving information from these collections. Examples include
Signature~\cite{signature}, Word Cruncher~\cite{wordcruncher}, Word
Smith Tools~\cite{wordsmithtools}, and Intext~\cite{intext}.  Such
tools tend not to be interactive.  Also, analyzing a multi-gigabyte
corpus tends to be slow.

\begin{figure}
 \center{
\includegraphics[width=1.40\columnwidth]{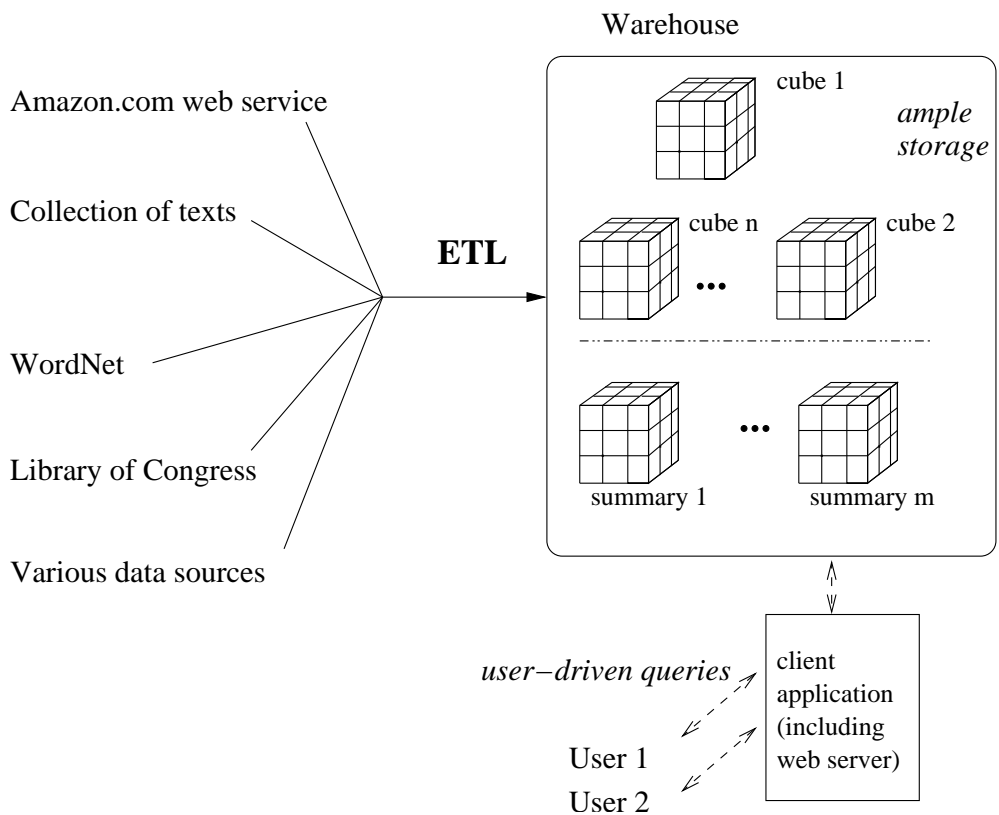}
 }
 \caption{\label{fig-wowdiagram} WoW architecture.}
 \end{figure}

We propose the creation of user-driven tools to interface with a
\emph{(Data) Warehouse of Words} (WoW) (see
Fig.~\ref{fig-wowdiagram}).  A WoW is built by an \emph{Extraction,
Transformation, and Loading} (ETL) procedure, which processes the text
and aggregates data from different sources.

\begin{figure}
 \center{
\includegraphics[width=0.80\columnwidth]{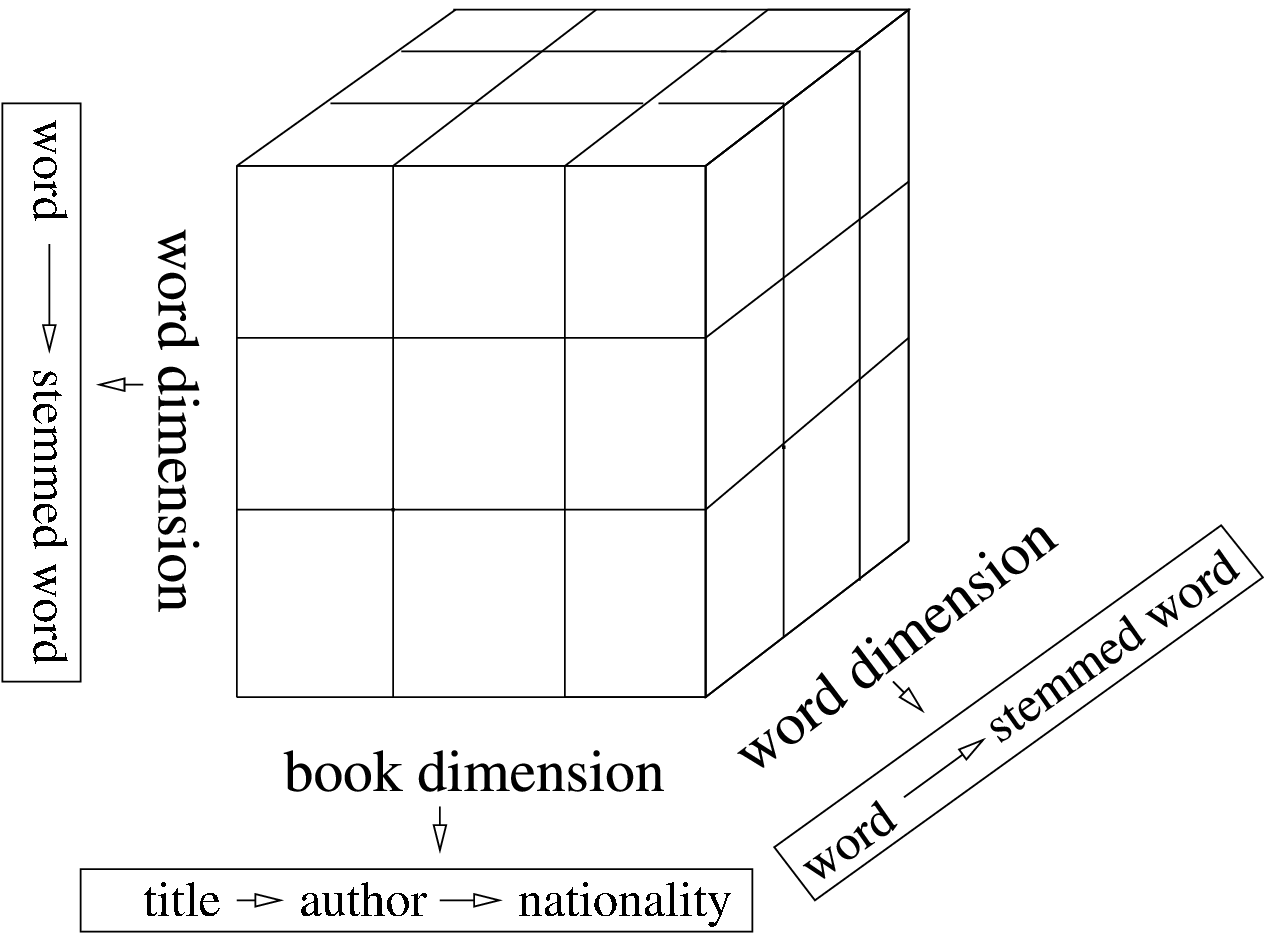}
 }
 \caption{\label{fig-simplecube} A simple data cube with three dimensions.}
 \end{figure}

A WoW stores its data in \emph{data cubes}~\cite{graycube}.  A data
cube can be abstracted as a $k$-dimensional array with several
predefined operations such as \emph{slicing, dicing, rolling up} and
\emph{drilling down}. These operations allow the user to focus on just
some subset of the data, at the desired granularity. While a data cube
may have 15 dimensions or more, the user may be only interested in 2
or 3 dimensions at a time.  See Fig.~\ref{fig-simplecube} for an
example of a 3-dimensional data cube with two word dimensions and a
book dimension.  An example cell might record a count of 10 for
(``cat'',``dog'') in \textit{Ivanhoe}, and this cube could be used to
study cooccurrences across several documents.  Moreover, the attribute
values of the example dimensions belong to a hierarchy: given the
title of a book, we can ``roll up'' to the author of the book and
finally to the author's nationality.

On-Line Analytical Processing (OLAP) provides near constant-time
answers to queries over large multidimensional data
sets~\cite{codd93}.  For example, a user may be interested in
comparing word or punctuation frequencies of two authors in the past
10 years.  In a standard relational database system this type of query
may be expensive.  OLAP, however, seeks to solve queries in a matter
of seconds before the user's train of thought has been lost.
Typically, fast results are at the expense of increased storage, by
precomputing \emph{summary} data cubes.

OLAP is especially applicable when many aggregate queries such as sum
and average are of interest.  Thus, data warehouses and OLAP have been
used widely in business applications.  Only recently have attempts
been made to handle scientific information in an OLAP
environment~\cite{882090}.  If we exclude Information Retrieval
(IR)~\cite{345656,moth:doccube}, this paper is the first attempt 
to apply OLAP to literature. OLAP has been used in
conjunction with text mining~\cite{Sullivan2001}, 
informetrics/bibliomining~\cite{niem:mddata-model-informetrics},
and to study literary titles~\cite{bern:a-juste-titre}, however.

The main advantage a user-driven OLAP tool would provide is
flexibility.  While IR and Artificial Intelligence tools are well
suited to their single function, a user-driven tool gives a wide
variety of users the freedom to pursue their individual research.

The end users of this OLAP system in most cases would be similar to
the existent business users in their lack of database query skills.
Most users would be unwilling to learn a multidimensional query
language like MDX~\cite{SQLServerMDX2005}.  Unlike business
executives, academics are usually unable to finance technical
assistants: they must be able to issue the queries themselves.  A
simple user-driven application is the most reasonable solution for
those users not already accustomed to writing their own MDX or SQL
queries.

Section 2 provides justification for this problem by presenting
several research areas where a user-driven tool would be of benefit.
Section 3 provides an overview of the ETL required to build the WoW.
Section 4 concludes with the schema of the WoW and a description of
queries the tool will support.  Though there currently exist literary
analysis tools, these types of applications do not take advantage of
the hierarchical structure of literary data.  Exploiting hierarchy is
a key notion in OLAP.

\section{Practical Applications}

User-driven analytical tools are used in the humanities for author
attribution, lexical analysis, and stylometric analysis.  We review
here some of the applications a WoW could support.

Author attribution is determining the authorship of an anonymous piece
of writing through various stylistic and statistical methods.
Mendenhall was the pioneer of this area with his study of word
lengths~\cite{Mendenhall}.  This field was made famous in the 1990's
by Foster's work~\cite{foster1,foster2} in attributing the authorship
of \textit{A Funeral Elegy} to Shakespeare.  Foster was also
responsible for determining the author of \textit{Primary Colors} and
has testified in a number of criminal court cases such as the trial of
Theodore Kaczynski (The Unabomber).  Though automatic author
attribution has been implemented with reasonable measures of
success~\cite{autoatt1,autoatt2,autoatt3}, the complexity of language
and stylistic analysis, as well as the fact that language is forever
evolving, makes it difficult to automate the process reliably over an
extended period of time.

Lexical analysis includes many measurements of vocabulary usage such
as \emph{Type-Token Ratio}, \emph{Number of Different Words} and
\emph{Mean Word Frequency}~\cite{vocabmeasure}.  These calculations
are highly aggregatable since they are applied to a single book, a
collection of books by a single author or time period, or an entire
collection of books. We can also study the relations between these
measures as is possible in a data cube.

Stylometric analysis not only considers the words in use but also
accounts for other statistical elements of style such as word length,
sentence length, use of punctuation and many other features.  Analysis
of this type tends to be used in assisting with author attribution as
well as studying the development of an author over
time~\cite{stylometry}. It has been shown that the frequencies and distributions
of words and recurrent phraseology can identify significant linguistic features
which literary critics may not see~\cite{StubbsConrad2005}.

Even analogies, and thus semantics, can be studied using a WoW data
cube.  Analogies of the form \emph{$A$ is to $B$ as $C$ is to
$D$}~\cite{Turney} can be characterized by cooccurrences: two words
connected by a \emph{joining word} such as $has$, $on$, and $with$ (64
joining words were initially proposed~\cite{Turney}).  Pairs of words
related by similar joining terms are analogous.

Other applications we foresee include user-driven mining for
frequent phrases~\cite{Blogpulse}, computer-assisted topos searches~\cite{SatorBase}, providing rich
Information Retrieval feedback, and even user-driven exploration in
order to improve computational linguistics algorithms.

\section{WoW Creation}

The development of our application begins with the creation of the
WoW.  This involves the three stages of ETL mentioned previously.  The
extraction in our case will involve the plain text and XML documents
of Project Gutenberg~\cite{Gutenberg}, a large corpus of literary
works that is not in a suitable form for immediate analysis (other
book collections might be added later).  Project Gutenberg's documents
contain irrelevant data such as the disclaimer and information on when
the document was created.  Also included in each preface are details
about the author, date of publication, and other facts which must be
extracted for indexing or statistical purposes.  The transformation
phase will involve the calculation of all data that will be stored in
the WoW such as word frequency, punctuation frequency, and sentence
lengths.  The loading phase will involve the actual creation and
storage of the data cubes containing the calculated items.

Our WoW is forced to deal with the same issues as any other data
warehouse.  At times data, such as the author's nationality, is
missing and must be handled.  Also, new books are added to corpora
daily, and a means for loading these new books into the WoW must be
created.

\section{WoW Schema}
The main strength of an OLAP application is its efficient evaluation
of aggregate queries across several dimensions and at different level
of granularity.  These queries generally take advantage of the
hierarchical nature of cube dimensions and the hierarchy is not always
obvious.  The schema of the WoW requires that the hierarchies for both
books and words be considered.

\subsection{Dimension Hierarchies}

\begin{figure}
\subfigure[Book hierarchy.]{
\label{bookhier}
\centering\includegraphics[width=0.80\columnwidth]{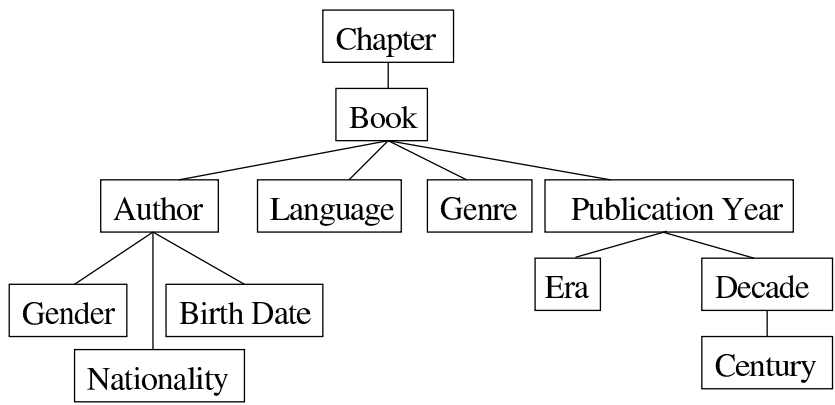}
}
\subfigure[Word hierarchy.]{
\label{wordhier}
\centering\includegraphics[width=0.80\columnwidth]{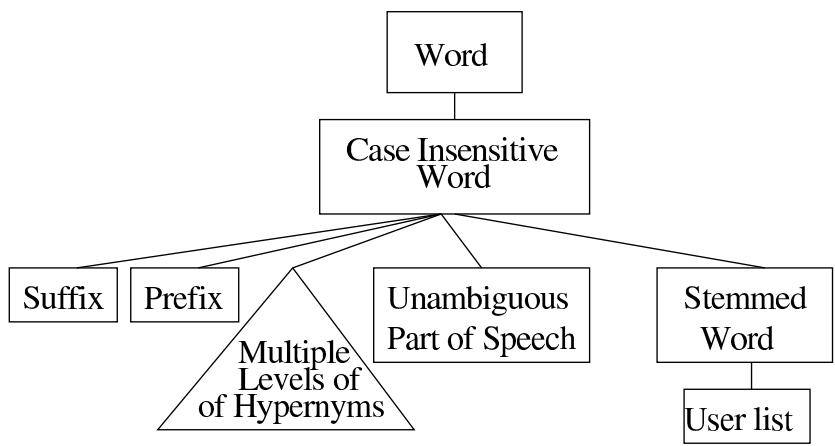}
}
\caption{Some WoW dimensions.}
\end{figure}

The ``book'' hierarchy maintains its finest detail at the level of
chapters and is shown in Figure~\ref{bookhier}.

The year of publication may be generalized to a literary era (eg
Victorian); alternatively, the year may be generalized to decade and
then to century.  Note that eras may not fit nicely into decades or
centuries.

Several natural generalizations may help word studies.  Refer to
Figure~\ref{wordhier}, where stemming groups together the forms
sharing a stem.  Alternatively, words can be grouped according to
their final suffix.  Some words (eg \textit{skit}) can be
unambiguously classified by part of speech, whereas classifying other
polysemous words may be impossible for our system (even if we try to
employ NLP parsers, which is not planned).  Thus ``unknown'' may be a
common generalization in the WoW.

Hypernyms, for instance as provided by WordNet or by the
classification in Roget's Thesaurus, provide another way to group
words.  This poses several difficulties involving polysemy whose
resolution is ongoing.

Finally, tools such as Signature allow user-specified word lists.
Given a set of ``interesting'' word stems, a stemmed word can be
classified as belonging to [one of] the user's lists or belonging to
no list\footnote{``Highly masculine'' words might help a feminist
literary analysis.}.

These hierarchies allow for \emph{rollup} queries (essentially
generalizations) to be evaluated.  Instead of finding the frequent
words used in a chapter or book, one might be interested in the
frequent words used by an author or used in a time period.  The
proposed WoW schema based on these hierarchies contains the following
data cubes.

\subsection{Cubes}

To support the initial stylometric, analogy, and phrase-use queries,
the WoW contains several cubes. We mention two.

\begin{enumerate}
\itemsep=0pt\topsep=0pt\partopsep=0pt
\parskip=0pt\parsep=0pt
\item
\textbf{Sentence Style} (Book $\times$ Word $\times$ WordCount
$\times$ CommaCount $\times$ ColonSemicolonCount $\times$
StopwordCount $\rightarrow$ Occurrence Count).  Each ``Count'' is an
integer, and the Word dimension represents the first word in a
sentence.  Many practical queries involve rollups of this cube.  For
instance, the average sentence length in each century can be computed
from this, or we could study the use of commas by authors who write
long sentences.

\item
\textbf{Short Phrase} (Book$\times$Word$\times$Word$\times$Word
$\times$Word $\rightarrow$ OccurrenceCount).  The cube records all
sequences of 4 words, and it could be used to explore common (or rare)
phrases by authors or time periods.

\end{enumerate}

These cubes will allow for many queries to be evaluated and would aid
in all of the previously mentioned practical applications as well as a
variety of other studies.  We believe the development of new and more
advanced queries will be stimulated by the creation of this system,
since there has yet to be a literary OLAP system taking such advantage
of hierarchies.

\bibliographystyle{alpha}
\bibliography{cline05,../litOLAP/litolap.bib}

\end{document}